\documentclass[12pt]{article}
\usepackage[dvips]{epsfig,graphics,color}
\begin{document}
\title{Similarity between Grover's quantum search algorithm and
classical two-body collisions}
\author{\small{} Jingfu Zhang, and Zhiheng Lu  \\
\small{}Department of Physics,\\
\small{}Beijing Normal University, Beijing, 100875, Peoples'
Republic of China\\}
\date{}
\maketitle

\begin{center}\bf Abstract\end{center}

 By studying the diffusion operator in Grover's quantum search
algorithm, we find a mathematical analogy between quantum
searching and classical, elastic two-body collisions. We exploit
this analogy using a thought experiment involving multiple
collisions between two-bodies to help illuminate why Grover's
quantum search algorithm works. Related issues are discussed.

\vspace*{0.4cm}

   Quantum computation is based on qubits. A qubit, literally a
 quantum bit of information, can be represented as a unit
 vector in a two-dimensional complex vector space. A typical vector
 can be written as $|\psi>=\alpha|0>+\beta|1>$, with
 $|\alpha|^{2}+|\beta|^{2}=1$. Measurement of a qubit yields
 $|\alpha|^{2}$ and $|\beta|^{2}$, the probabilities of the system
 being in states $|0>$ and $|1>$. Unlike a classical bit, a qubit
 represents a superposition of the
two states. This property leads to quantum parallel computation,
which has been widely used in quantum algorithms [1]. Grover
proposed a quantum search algorithm which can realize fast
searching, such as finding an object in unsorted data consisting
of $N$ items [2]. This algorithm can be described as follows.
Initially, an $n$ qubit system is set in an equal superposition of
all basis states expressed as
\begin{equation}\label{1}
  |\Psi_{0}>={1\over \sqrt{N}}\sum_{i=0}^{N-1}|i>,
\end{equation}
where the set $\{|i>\}$ is orthonormal, and $N$ is usually far
greater than 1. Each basis state corresponds to an item in the
data. The problem is how to find the very basis state that
corresponds to the object. This particular state is defined as the
marked state, and the other states are defined as the collective
state [3]. $C$ and $D$ are used to denote suitable inversion and
diffusion operators, respectively. If $C$ is applied to a
superposition of states, it only inverts (i.e., changes the
algebraic sign of) the amplitude in the marked state, and leaves
the states in the collective state unaltered. $D$ is defined as
$D\equiv 2P-I$, where $P\equiv (1/N) \sum_{i,j=0}^{N-1}|i><j|$. In
matrix notation, $P$ is an $N\times N$ matrix whose entries are
all $1/N$, and $I$ is an $N\times N$ unit matrix. A compound
operator is defined as $U\equiv DC$. Each operation of $U$ is
called an iteration. After $U$ is repeated $O(\sqrt{N})$ times,
the amplitude in the marked state can approach 1. When a
measurement is made, the probability of getting this state
approaches 1 [2]. The algorithm can be generalized to the case
where the marked state contains multiple basis states.[4] The
search algorithm has become a hot topic in quantum information
[5]-[7]. Grover outlined the steps that led to his algorithm[8],
and proposed a classical analog of quantum search using a coupled
pendulum model[9]. R. Josza provided another path toward
understanding the nature of Grover's algorithm[10]. In this paper,
we discuss the similarity between quantum searching and classical
two-body collisions and some related problems such as the limits
on the algorithm. The similarity helps to understand why the
algorithm works.

We now examine some properties of $D$. If $D$ is applied to a
system in state $\sum_{i=0}^{N-1}c_{i}|i>$, it transforms the
system into state  $ \sum_{i=0}^{N-1}d_{i}|i> $. The relation
between $c_{i}$ and $d_{i}$ is

\begin{equation}\label{2}
 d_{i}=2A-c_{i},
\end{equation}
where $A$ is the average of all amplitudes of the system, namely,
$A=(1/N)\sum_{i=0}^{N-1}c_{i}$. Through easy computations, we find
that

\begin{equation}\label{3}
  \sum_{i=0}^{N-1}d_{i}=\sum_{i=0}^{N-1}c_{i}
\end{equation}
and

\begin{equation}\label{4}
  \sum_{i=0}^{N-1}d_{i}d_{i}^{*}=\sum_{i=0}^{N-1}c_{i}c_{i}^{*},
\end{equation}
where $c_{i}^{*}$ denotes the complex conjugation of $ c_{i}$.
Eq.(3) shows that the transformation by $D$ preserves the sum of
all amplitudes. Eq.(4) results from normalization of the wave
function.

We suppose that the collective state contains $N_{1}$ basis states
with equal amplitude $a_{0}$ and the marked state contains $N_{2}$
basis states with equal amplitude $b_{0}$ before $U$ is applied.
From now on, we assume that all amplitudes are real numbers. $C$
transforms $b_{0}$ to $-b_{0}$, and leaves $a_{0}$ unaltered; $D$
transforms $-b_{0}$ to $b_{1}$, and $a_{0}$ to $a_{1}$. According
to Eqs. (3) and (4), two equations

\begin{equation}\label{5}
N_{1}a_{1}+N_{2}b_{1}=N_{1}a_{0}-N_{2}b_{0},
\end{equation}

\begin{equation}\label{6}
N_{1}a_{1}^{2}+N_{2}b_{1}^{2}=N_{1}a_{0}^{2}+N_{2}b_{0}^{2}
\end{equation}
are obtained. They remind us that the operation of $D$ is
analogous to the elastic collision of two bodies. We assume they
are two rigid balls with masses $m_{1}$ and $m_{2}$. Ball 1 and
ball 2 have velocities $u$ and $v$, respectively. Before the first
collision, these velocities are indicated by $u_{0}$ and $-v_{0}$;
after the collision, they are denoted by $u_{1}$ and $v_{1}$. All
velocities are confined to one straight line and the minus sign
means that the two balls move in opposite direction before the
collision. We obtain two equations
\begin{equation}\label{7}
m_{1}u_{1}+m_{2}v_{1}=m_{1}u_{0}-m_{2}v_{0},
\end{equation}

\begin{equation}\label{8}
{1\over2}m_{1}u_{1}^{2}+{1\over2}m_{2}v_{1}^{2}={1\over2}m_{1}u_{0}^{2}+{1\over
2}m_{2}v_{0}^{2}
\end{equation}
based on the conservation of momentum and mechanical energy [11].
We find the forms of Eqs. (7)-(8) similar to the forms of Eqs.
(5)-(6). Their solutions also have similar forms. The solutions
for Eqs. (7) and (8) are
\begin{equation}\label{9}
u_{1}={(m_{1}-m_{2})u_{0}-2m_{2}v_{0}\over m_{1}+m_{2}},
\end{equation}
\begin{equation}\label{10}
v_{1}={2m_{1}u_{0}+(m_{1}-m_{2})v_{0}\over m_{1}+m_{2}}.
\end{equation}
Let $v_{c}=(m_{1}u_{0}-m_{2}v_{0})/(m_{1}+m_{2})$ be the velocity
of the mass center of the system. $v_{c}$ is constant during the
collision of the two balls. $u_{1}$ and $v_{1}$ can thus be
represented as
\begin{equation}\label{11}
  u_{1}=2v_{c}-u_{0}
\end{equation}

\begin{equation}\label{12}
  v_{1}=2v_{c}+v_{0}
\end{equation}

  Now, we assume that ball 1 and ball 2 consist of
 $N_{1}$ and $N_{2}$ particles respectively. For convenience, we
denote these particles as $p_{i}(i=0,1,...,N-1)$. The mass of
 each $p_{i}$ is denoted by $M_{i}=m_{0}$, where $m_{0}$ denotes unit
mass. Therefore $m_{1}=N_{1}m_{0}$, $m_{2}=N_{2}m_{0}$, and
$N_{1}+N_{2}=N$. Because of the similarity between Eqs. (7)-(8)
and Eqs. (5)-(6), a one-to-one correspondence exists between the
amplitude of $|i>$ and the velocity of $p_{i}$. All velocities are
confined to one straight line because all amplitudes are assumed
to be real numbers. The probability of getting $|i>$ is analogous
to the kinetic energy of $p_{i}$. $N_{1}$ particles in ball 1 and
$N_{2}$ particles in ball 2 are analogous to $N_{1}$ basis states
in the collective state and $N_{2}$ basis states in the marked
state, respectively. Because $v_{c}$ is also represented as
$v_{c}=(1/N)\sum_{i=0}^{N-1}V_{i}$, where $V_{i}$ is the velocity
of $p_{i}$, it can be concluded that $v_{c}$ is analogous to $A$
in Eq.(2). We summarize the above-mentioned correspondences in
 Table $\rm 1$.

\begin{center}
Table $\rm 1$: Correspondences between Grover's search algorithm
and classical two-body collisions
\begin{tabular}{|c|c|}
\hline
 Grover's search algorithm   & classical two-body collisions   \\
 \hline
 Eqs. (5)-(6)& Eqs. (7)-(8)\\
 \hline
state $|i>$ & particle $p_{i}$\\
\hline
 amplitude of $|i>$ & velocity of $p_{i}$\\
 \hline
 probability of getting $|i>$ & kinetic energy of $p_{i}$ \\
 \hline
marked state & ball 2 \\
\hline
collective state & ball 1\\
\hline
 A ( average of amplitudes) & $v_{c}$ (center of
 mass  velocity )\\
\hline
\end{tabular}
\end{center}
\vspace*{0.4cm}

 Initially, two balls move with the same velocity, i.e.,
all particles move with the same velocity. This is analogous to
the initial equal superposition of states in quantum searching.
The operation of $C$ is analogous to the operation that inverts
the velocities of $N_{2}$ particles in ball 2, and leaves
particles in ball 1 unaltered. From Eqs. (11) and (12), in the
center of mass frame, we find that before the collision, two balls
have velocities $u_{0}-v_{c}$ and $-v_{0}-v_{c}$; after the
collision, their velocities are transformed to $v_{c}-u_{0}$ and
$v_{c}+v_{0}$.
 This fact means that the velocities are inverted by
the collision. In the reference frame of the laboratory, however,
the after-collision velocities are $2v_{c}-u_{0}$ and
$2v_{c}+v_{0}$, respectively. Application of the diffusion
operator $D$, which is also called inversion about average,
transforms the amplitudes of the collective state and the marked
state from $a_{0}$ and $-b_{0}$ to $2A-a_{0}$ and $2A+b_{0}$.
Therefore, the similarity between operation of $D$ and the
collision is obvious.

  According to the classical analog of the quantum searching algorithm,
we conceive a thought experiment to simulate the algorithm. Ball 1
and ball 2 move rightwards along a horizontal and straight path
without friction with the same initial velocity $u_{0}=v_{0}=v$ as
shown in Fig.1a.
\begin{figure}[ht]
\centerline{\epsfig{figure=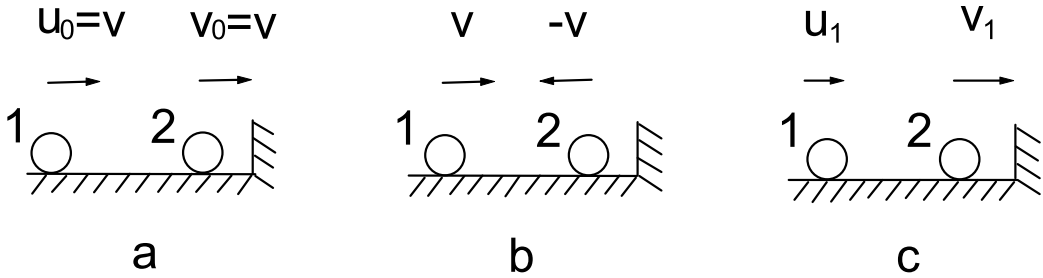,width=6in}}
\begin{center}
 \begin{minipage}{120mm}
 \caption{{\bf A scheme used to simulate the quantum search algorithm using
classical collisions}
 The two balls of masses $m_{1}$ and $m_{2}$ are
 moving with the same velocity along a horizontal and straight path without
 friction (Fig.1a).
 After ball 2 collides with the obstacle at the right end of the path,
 its velocity is inverted (Fig.1b). After the collision
 of the two balls, the velocities become $u_{1}$ and $v_{1}$
  (Fig.1c).}
\end{minipage}
\end{center}
 \label{1}
\end{figure}
In order to invert the velocity, we set an obstacle at the right
end of the path. The velocity of ball 2 is inverted but its
modulus remains unaltered after it collides with the obstacle. The
velocities of the two balls after the collision are $v$ and $-v$
as shown in Fig.1b. This collision is analogous to the operation
of $C$ used in the quantum searching algorithm. After ball 2
collides with the obstacle, it collides with ball 1. The second
collision is analogous to operation by $D$. The velocities after
this collision are $u_{1}$ and $v_{1}$ as shown in Fig.1c. The two
collisions that are analogous to $C$ and $D$ constitute the
complete iteration analogous to $U$ used in the quantum searching.
There are 3 possible cases after the collision of the two balls:
1) The two balls both move rightwards, 2) ball 1 moves leftwards
and ball 2 moves rightwards, and 3) the two balls both move
leftwards. In some cases, the next collision of two balls cannot
happen after the velocity of ball 2 is inverted again. Because we
only are concerned with collisions and have no interest in the
positions of the two balls, we can exchange the two balls'
positions or set an obstacle at the left end of the path in order
to make the iteration continue. Assuming the two balls have
velocities $u_{n}$ and $v_{n}$ after $n$ iterations, we obtain the
recursion equations in matrix notation
\begin{equation}\label{13}
\left(\begin{array}{cc}
     u_{n+1} \\
      v_{n+1}
\end{array}\right)=
\left(\begin{array}{cc}
     {N_{1}-N_{2}}\over N_{1}+N_{2} & {-2N_{2}}\over N_{1}+N_{2} \\
     {2N_{1}}\over N_{1}+N_{2} & {N_{1}-N_{2}}\over N_{1}+N_{2}
\end{array}\right)
\left(\begin{array}{c}
    u_{n} \\
    v_{n} \
\end{array}\right),
\end{equation}
by replacing the subscripts 1 and 0 in Eqs.(9) and (10) by $n+1$
and $n$, respectively, and by using the conditions that
$m_{1}=N_{1}m_{0}$, and $m_{2}=N_{2}m_{0}$. For the quantum
system, after $U$ is repeated $n$ times, the amplitudes in the
collective state and marked state are transformed to $a_{n}$ and
$b_{n}$. One can directly obtain the recursion equations
\begin{equation}\label{14}
\left(\begin{array}{c}
  a_{n+1} \\
 b_{n+1}
\end{array}\right)=
\left(\begin{array}{cc}
 {N_{1}-N_{2}}\over{ N_{1}+N_{2}} & {-2N_{2}}\over{ N_{1}+N_{2}} \\
  {2N_{1}}\over{ N_{1}+N_{2}} & {N_{1}-N_{2}}\over{ N_{1}+N_{2}}
\end{array}\right)
\left(\begin{array}{c}
  a_{n} \\
  b_{n}
\end{array}\right)
\end{equation}
from Eq.(13), based on the above-mentioned similarities. The
initial conditions are $u_{0}=v_{0}=v$ for Eq.(13), and
$a_{0}=b_{0}=1/\sqrt{N}$ for Eq.(14). The $2\times 2$ matrix in
Eq.(14) is the product of Grover's diffusion operator and suitable
inversion operator, and is denoted by
\begin{equation}\label{15}
 T=\left(\begin{array}{cc}
 {N_{1}-N_{2}}\over{ N_{1}+N_{2}} & {-2N_{2}}\over{ N_{1}+N_{2}} \\
  {2N_{1}}\over{ N_{1}+N_{2}} & {N_{1}-N_{2}}\over{ N_{1}+N_{2}}
\end{array}\right).
\end{equation}

  We will proceed to solve Eqs.(13) and (14) using linear algebra
[12]. One finds that
\begin{equation}\label{16}
\left(\begin{array}{cc}
     u_{n} \\
      v_{n}
\end{array}\right)=T^{n}\left(\begin{array}{cc}
     u_{0} \\
      v_{0}
\end{array}\right),
\end{equation}
\begin{equation}\label{17}
\left(\begin{array}{c}
  a_{n} \\
 b_{n}
\end{array}\right)=T^{n}\left(\begin{array}{c}
  a_{0} \\
 b_{0}
\end{array}\right),
\end{equation}
where $n=1,2,\cdots$. In order to obtain explicit expressions for
$u_{n}$, $v_{n}$, $a_{n}$ and $b_{n}$, we transform $T$ to the
diagonal matrix
\begin{equation}\label{18}
  T_{D}=S^{-1}TS\equiv\left(\begin{array}{cc}
 \lambda_{+} & 0 \\
  0 & \lambda_{-}
\end{array}\right).
\end{equation}
The eigenvalues $\lambda_{\pm}$ of the matrix $T$ are the
solutions of $\det(T-\lambda I)=0$. They can be expressed as
\begin{equation}\label{19}
\lambda_{\pm}=\begin{array}{c}
  {N_{1}-N_{2}\pm 2i\sqrt{N_{1}N_{2}}}\over{N_{1}+N_{2}}
\end{array}.
\end{equation}
$S$ and $S^{-1}$ are expressed as
\begin{equation}\label{20}
 S= \left(\begin{array}{cc}
 1 & 1 \\
  -i\sqrt{{N_{1}}\over{N_{2}}} & i\sqrt{{N_{1}}\over{N_{2}}}
\end{array}\right),
\end{equation}
\begin{equation}\label{21}
 S^{-1}= \frac{1}{2}\left(\begin{array}{cc}
 1 & i\sqrt{{N_{2}}\over{N_{1}}} \\
  1 & -i\sqrt{{N_{2}}\over{N_{1}}}
\end{array}\right).
\end{equation}
One can easily find that
\begin{equation}\label{22}
  T^{n}=ST_{D}^{n}S^{-1}.
\end{equation}
If $N_{2}=1$, and $N\gg 1$, $T_{D}$ can be expressed as
\begin{equation}\label{23}
  T_{D}=\left(\begin{array}{cc}
 e^{i2\theta} & 0 \\
  0 & e^{-i2\theta}
\end{array}\right),
\end{equation}
using $\tan 2\theta \approx 2\theta$, where
$\theta\equiv\frac{1}{\sqrt{N}}$. Using Eqs.(20)-(23), one can
derive that
\begin{equation}\label{24}
  T^{n}=\left(\begin{array}{cc}
 \cos(2n\theta) & -\frac{1}{\sqrt{N-1}}\sin(2n\theta)\\
  \sqrt{N-1}\sin(2n\theta)& \cos(2n\theta)
\end{array}\right).
\end{equation}
Using Eqs.(16), (17) and (24), we obtain the explicit expressions
\begin{equation}\label{25}
\left(\begin{array}{cc}
     u_{n} \\
      v_{n}
\end{array}\right)=
\left(\begin{array}{c}
     v\frac{\sqrt{N}}{\sqrt{N-1}}cos((2n+1)\theta)\\
    v\sqrt{N}sin((2n+1)\theta)
\end{array}\right),
\end{equation}

\begin{equation}\label{26}
\left(\begin{array}{cc}
    a_{n} \\
      b_{n}
\end{array}\right)=
\left(\begin{array}{c}
     \frac{1}{\sqrt{N-1}}cos((2n+1)\theta)\\
    sin((2n+1)\theta)
\end{array}\right),
\end{equation}
noting that $\sin\theta\approx\frac{1}{\sqrt{N}}$, and $\cos\theta
\approx \sqrt{\frac{N-1}{N}}$. During the course of collisions,
energy is transferred between the two balls through the change of
their velocities. Let $n_{0}\approx (\pi\sqrt{N}/4-1/2)$ be the
integer obtained by rounding $(\pi\sqrt{N}/4-1/2)$. When
$n=n_{0}$, ball 2 can acquire almost all the mechanical energy of
the system. Correspondingly, for the quantum searching algorithm,
the amplitude in the marked state can approach 1 if the number of
repetitions of $U$ is $n_{0}$. The quantum system lies in the
marked state if a measurement is made. Our results show agreements
with Refs.[4][13][14]. We now discuss some related problems based
on these discussed similarities.

{\bf1.Simulating Grover quantum searching algorithm on a classical
system} Because $p_{i}$ is analogous to $|i>$, a system consisting
of $2^{n}$ particles $p_{i}(i=0,1,2,...,2^{n}-1)$ can be used to
simulate a quantum system that stores $n$ qubits of information.
For example, if $m_{1}=3m_{0}$ and $m_{2}=m_{0}$, the classical
system can simulate a two qubit system with $N_{1}=3$, $N_{2}=1$,
and $N=4$. After exactly one iteration, we find that $u_{1}=0$,
$v_{1}=2v$. This result means that ball 2 obtains all the energy
of the system, which is analogous to the probability equal to 1 of
getting the marked state. Our result is in agreement with the
result demonstrated on an NMR quantum computer [5]. If
$m_{1}=7m_{0}$, and $m_{2}=m_{0}$, the following results are
obtained:
\begin{equation}\label{27}
\left(\begin{array}{c}
  u_{1} \\
  v_{1}
\end{array}\right)
=\left(\begin{array}{c}
  1/2 \\
  5/2
\end{array}\right)v,
\end{equation}

\begin{equation}\label{28}
\left(\begin{array}{c}
  u_{2} \\
  v_{2}
\end{array}\right)=\left(\begin{array}{c}
  -1/4 \\
  11/4
\end{array}\right)v.
\end{equation}
Ball 2 nearly gets the entire energy of the system after 2
iterations. Because mass is continuous, the case where
$m_{2}=m_{0}$, and $m_{1}=N_{1}m_{0}$ is equivalent to the case
where $m_{2}=km_{0}$, and $m_{1}=kN_{1}m_{0}$ ($k$ is a positive
integer). The case where $m_{1}=3m_{0}$, and $m_{2}=m_{0}$ is
analogous to the case where $N_{2}=N/4$ as discussed in Ref. [4],
in which after one iteration, the probability of getting the
marked state is 1. If we choose the condition that
$m_{1}/m_{2}=p/q$, where $q\neq 1$, and $p$ and $q$ are positive
integers with no common factor, we can simulate the quantum search
algorithm in the case where the marked state contains multiple
basis states, which cannot be realized by NMR [15].

{\bf2. Limits on quantum searching} Because the two balls exchange
their energy after they collide with each other, neither of them
can get all the energy of the system if $m_{1}=m_{2}$. This case
is analogous to the case where the search algorithm is invalid,
more specifically, where $N_{1}=N_{2}$ [16]. We also find that
\begin{equation}\label{29}
\left(\begin{array}{c}
    u_{1} \\
    v_{1} \
  \end{array}\right)=
  \left(\begin{array}{c}
    1-4N_{2}/N \\
    3-4N_{2}/N \
  \end{array}\right)v,
\end{equation}
using the initial condition $u_{0}=v_{0}=v$. Eq.(29) means that if
$N_{2}>N/4$, the velocity of ball 1 is inverted after the first
iteration. It is possible that the two balls move in the same
direction after the velocity of ball 2 is inverted in the second
iteration. The energy of ball 2 is reduced after it catches up
with ball 1 and collides with it. In this case, the searching
algorithm is not efficient [16].

{\bf3.Qualitative discussion of the needed number of iterations}
Considering the case where $m_{2}=m_{0}$, $m_{1}=(N-1)m_{0}$. The
initial condition is $u_{0}=v_{0}=v$, and $N \gg 1$. We obtain
$u_{1}\approx v$, and $v_{1}=3v $ after the first iteration. This
means that the velocity of ball 2 increases by $2v$ and the
velocity of ball 1 hardly change in any iteration. The velocity of
ball 2 increases by $2nv$ after $n$ iterations. When
$n\approx\sqrt{N}/2$, ball 2 gets almost all energy of the system.

 {\bf Acknowledgement} This work is supported by the National Nature Science
Foundation of China. We are also grateful to Professor Shouyong
Pei of Beijing Normal University for his helpful discussions on
the principle of quantum algorithm.

\newpage
\bibliographystyle{article}

\end{document}